\documentclass[traditabstract]{aa} 
\usepackage{graphicx,natbib,longtable,lscape,color,amssymb}
\newcommand{\FR}{FRI{\sl{CAT}}}
\newcommand{\FRo}{FR0{\sl{CAT}}}
\newcommand{\WHz}{\>{\rm W}\,{\rm Hz}^{-1}}

\begin{document}
\title{The LOFAR view of FR~0 radio galaxies}
\author{A. Capetti\inst{1} \and M. Brienza\inst{2,3} \and
  R.~D. Baldi\inst{4,5,6} \and G. Giovannini\inst{2,3} \and
  R. Morganti\inst{7,8} \and M.J. Hardcastle\inst{9} \and
  H.J.A. Rottgering\inst{10} \and G.F. Brunetti\inst{3} \and
  P.N. Best\inst{11} \and G. Miley\inst{10}}\institute{INAF -
  Osservatorio Astrofisico di Torino, Strada Osservatorio 20, I-10025
  Pino Torinese, Italy \and Dipartimento di Fisica e Astronomia,
  Universit\`a di Bologna, Via P. Gobetti 93/2, I-40129, Bologna,
  Italy \and INAF - Istituto di Radio Astronomia, Via P. Gobetti 101,
  I-40129 Bologna, Italy \and Department of Physics \& Astronomy,
  University of Southampton, Hampshire SO17 1BJ, Southampton \and
  Dipartimento di Fisica, Universit\'a degli Studi di Torino, via
  Pietro Giuria 1, 10125 Torino, Italy \and INAF - Istituto di
  Astrofisica e Planetologia Spaziali, via Fosso del Cavaliere 100,
  I-00133 Roma, Italy \and ASTRON, the Netherlands Institute of Radio
  Astronomy, Postbus 2, NL-7990 AA, Dwingeloo, the Netherlands \and
  Kapteyn Astronomical Institute, University of Groningen, PO Box 800,
  NL-9700 AV Groningen, the Netherlands \and Centre for Astrophysics
  Research, University of Hertfordshire, College Lane, Hatfield AL10
  9AB, UK \and Leiden Observatory, Leiden University, PO Box 9513,
  2300 RA Leiden, The Netherlands \and SUPA, Institute for Astronomy,
  Royal Observatory, Blackford Hill, Edinburgh, EH9 3HJ, UK} \date{}

\abstract{We explore the low-frequency radio properties of the sources
  in the Fanaroff-Riley class 0 catalog (FR0{\sl{CAT}}) as seen by the
  Low-Frequency ARray (LOFAR) observations at 150 MHz. This sample
  includes 104 compact radio active galactic nuclei (AGN) associated
  with nearby ($z<0.05$) massive early-type galaxies. Sixty-six
  FR0{\sl{CAT}} sources are in the sky regions observed by LOFAR and
  all of them are detected, usually showing point-like structures with
  sizes of $\lesssim$3-6 kpc. However, 12 FR~0s present resolved
  emission of low surface brightness, which contributes between 5\%
  and 40\% of the total radio power at 150 MHz, usually with a jetted
  morphology extending between 15 and 50 kpc. No extended emission is
  detected around the other FR~0s, with a typical luminosity limit of
  $\lesssim 5 \times 10^{22}$ W Hz$^{-1}$ over an area of 100 kpc
  $\times$ 100 kpc. The spectral slopes of FR~0s between 150 MHz and
  1.4 GHz span a broad range ($-0.7 \lesssim \alpha \lesssim 0.8$)
  with a median value of $\overline\alpha \sim 0.1$; 20\% of them have
  a steep spectrum ($\alpha \gtrsim 0.5$), which is an indication of
  the presence of substantial extended emission confined within the
  spatial resolution limit. The fraction of FR~0s showing evidence for
  the presence of jets, by including both spectral and morphological
  information, is at least $\sim 40\%$.  This study confirms that
  FR~0s and FR~Is can be interpreted as two extremes of a continuous
  population of jetted sources, with the FR~0s representing the low
  end in size and radio power.}

\maketitle

\section{Introduction}
\label{intro}

The majority of the radio-active galactic nuclei associated with low redshift
galaxies detected in recent surveys at 1.4 GHz (for example,
\citealt{best12}) are compact, with corresponding linear sizes
$\lesssim$ 10 kiloparsecs \citep{baldi09}. Earlier surveys (performed at a lower
frequency and sensitivity) were instead dominated by sources
extending over scales of hundreds of kpc (see 
\citealt{hardcastle98}). The ``compact'' sources were named ``FR~0s''
\citep{ghisellini11,sadler14,baldi15} as a convenient way to include
them into the canonical \citet{fanaroff74} classification scheme of
radio galaxies (RGs), referring to their lack of extended radio
emission. The information available from observations of FR~0s is
generally very limited, even in the radio band.  It is then still
unclear as to the nature of these sources and how they are related
to the other classes of RGs.

\citet{baldi18} selected a sample of compact radio sources named \FRo\
in order to perform a systematic study of FR~0s. \FRo\ is formed by
compact radio sources with a redshift $\leq 0.05$ selected by combining
observations from the National Radio Astronomy Observatory Very Large
Array Sky Survey (NVSS; \citealt{condon98}), the Faint Images of the
Radio Sky at Twenty centimeters survey, (FIRST,
\citealt{becker95,helfand15}), and the Sloan Digital Sky Survey (SDSS;
\citealt{york00b}). The \FRo\ selection is limited to the galaxies in
which various methods (see \citealt{best12} for details) indicate that
the emission is due to a radio-AGN, thus excluding star-forming
galaxies. \citeauthor{baldi18} included the sources that are brighter
than 5 mJy and with a limit to their angular size of 4$\arcsec$ in the
FIRST images for a total of 104 sources. Their luminosities at 1.4 GHz
are in the range $10^{22} \lesssim L_{1.4 {\rm GHz}} \lesssim 10^{24}
\WHz$.

The key question about the FR~0s pertains to why they do not show the
extended radio emission that characterizes the other classes of RGs:
and, additionally, whether this is due to different properties of
their central engines. Are FR~0s not able to produce relativistic
jets or is there an evolutionary link between them, that is to say,
whether FR Os are an early stage of FR Is.

While the radio structure of FR~0s and FR~Is is different, the nuclear
and host galaxy properties of these two classes are very similar
\citep{baldi18,torresi18}. There is only a difference in the host
optical luminosity of the two classes, those of the FR~Is being $\sim
60\%$ brighter that those of the FR~0s, but with a large superposition
of the two distributions.

Based on the relative number density of the two classes
\citeauthor{baldi18} were able to discard the scenario in which all
FR~0s are young RGs which will eventually evolve into extended radio
sources. Nonetheless, there must be an intermediate evolutionary stage
between the initial nuclear activity of FR~Is and FR~IIs before they
reach their typical linear sizes of hundreds of kpc: some FR~0s are
expected to be small because they are young. \citet{capetti19}
estimated that the fraction of FR~0s with a high curvature convex
spectrum, typical of young radio galaxies, is at most
$\sim$15\%.

FR~0s might instead be recurrent sources, characterized by short
phases of activity. Alternatively, \citet{baldi15} suggested that what
distinguished FR~0s from FR~Is is the lower-bulk Lorentz factor of the
jets in FR~0s, which are disrupted before they can emerge from the
host galaxies. Constraints on the jet speed and structure in FR~0s
might be obtained from high resolution imaging. The very long baseline
observations analyzed by \citet{cheng18} suggest the presence of
jetted structure in all but one of the sources of the subsample of 14
bright (flux densities $>$50 mJy) FR~0s they studied. Their pc-scale
structures indicate the presence of a broad range of relativistic
beaming factors.

\citet{capetti20} explored the large-scale environment of FR~0s. They
found that FR~0s are located in regions with an average number of
galaxies that is lower by a factor two with respect to FR~I. This
difference is driven by the large fraction (63\%) of FR~0s that are
located in groups formed of fewer than 15 galaxies. FR~Is rarely
(17\%) inhabit an environment like this. One possibility to account
for the connection between environment and the properties of the
extended radio emission is related to the stronger adiabatic losses of
the radio-emitting plasma (for example, \citealt{longair94}) in the poorer
environment of the FR~0s. However, \citet{baldi19} showed that most
FR~0s are still confined well within the core of the hot corona of
their host and the similarity of FR~0 and FR~I host galaxies suggests
that their coronae also have similar distribution of hot gas. The
differences in environment between FR~0s and FR~Is can instead be due
to an evolutionary link between local galaxy density, black hole spin
\citep{garofalo19}, jet power, and extended radio emission. In
addition to the lack of substantial extended radio emission that
defines the FR~0 class, the properties of the large scale environment
represent the first significant difference between these two
populations of low-power radio galaxies.

In this context, low-frequency and high sensitivity radio observations
of FR~0s might be used to address the following questions:

(1) Do FR~0s show low-frequency extended emission? If FR~0s are indeed
recurrent sources, one might expect to detect relic emission from a
previous activity phase, best observable at MHz-frequencies due to its
steep spectrum.

(2) What is the low-frequency spectral shape of FR~0s? Observations at
high resolution, required to spatially isolate any small-scale
extended emission, are only available for a minority of FR~0s. The
spectral index information can be used to infer the fraction of
optically thin, hence extended, emission present in FR~0s overcoming
the limited spatial resolution. The large frequency leverage opened by
the LOFAR \citep{vanhaarlem13} images is optimally suited to perform
this study.

We already studied the low-frequency properties of FR~0 radio galaxies
\citep{capetti19} by using the Alternative Data Release of the TIFR
GMRT Sky Survey (TGSS, GMRT; \citealt{swarup91,intema17}). This
analysis was however limited to a flux density limit of 17.5 mJy and
it returned an association for only 37 out of 104 FR~0 of the
\FRo\ sample. Nonetheless it was possible to conclude that 1) most
FR~0s (92\%) have a flat or inverted spectral shape ($\alpha <
0.5$)\footnote{Spectral indices $\alpha$ are defined as
  $F_{\nu}\propto\,\nu^{-\alpha}$.}  between 150 MHz and 1.4 GHz and
2) no extended emission is detected around them, corresponding to a
luminosity limit of $\lesssim 4 \times 10^{23}$ W Hz$^{-1}$.  The
observations that are being obtained with LOFAR, thanks to their
higher depth and resolution, can be used to improve significantly our
knowledge of the low-frequency radio properties of FR~0s.

The paper is organized as follows: in Sect. 2 we describe the LOFAR
observations available for the \FRo\ sample.  The results are
presented in Sect. 3 and discussed in Sect. 4. In Sect. 5 we draw our
conclusions.

\begin{figure*}
\includegraphics[scale=0.38,angle=0]{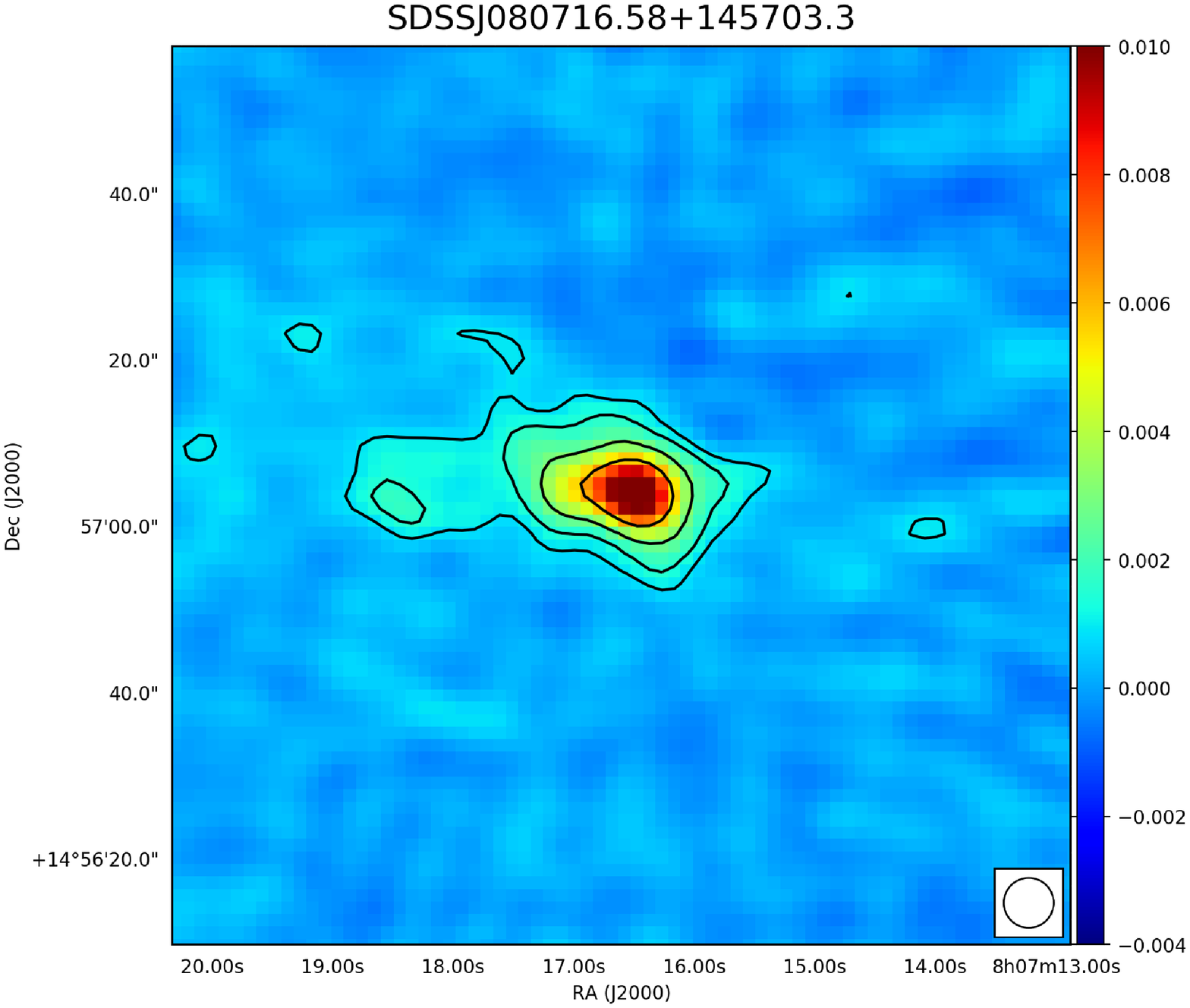}
\includegraphics[scale=0.38,angle=0]{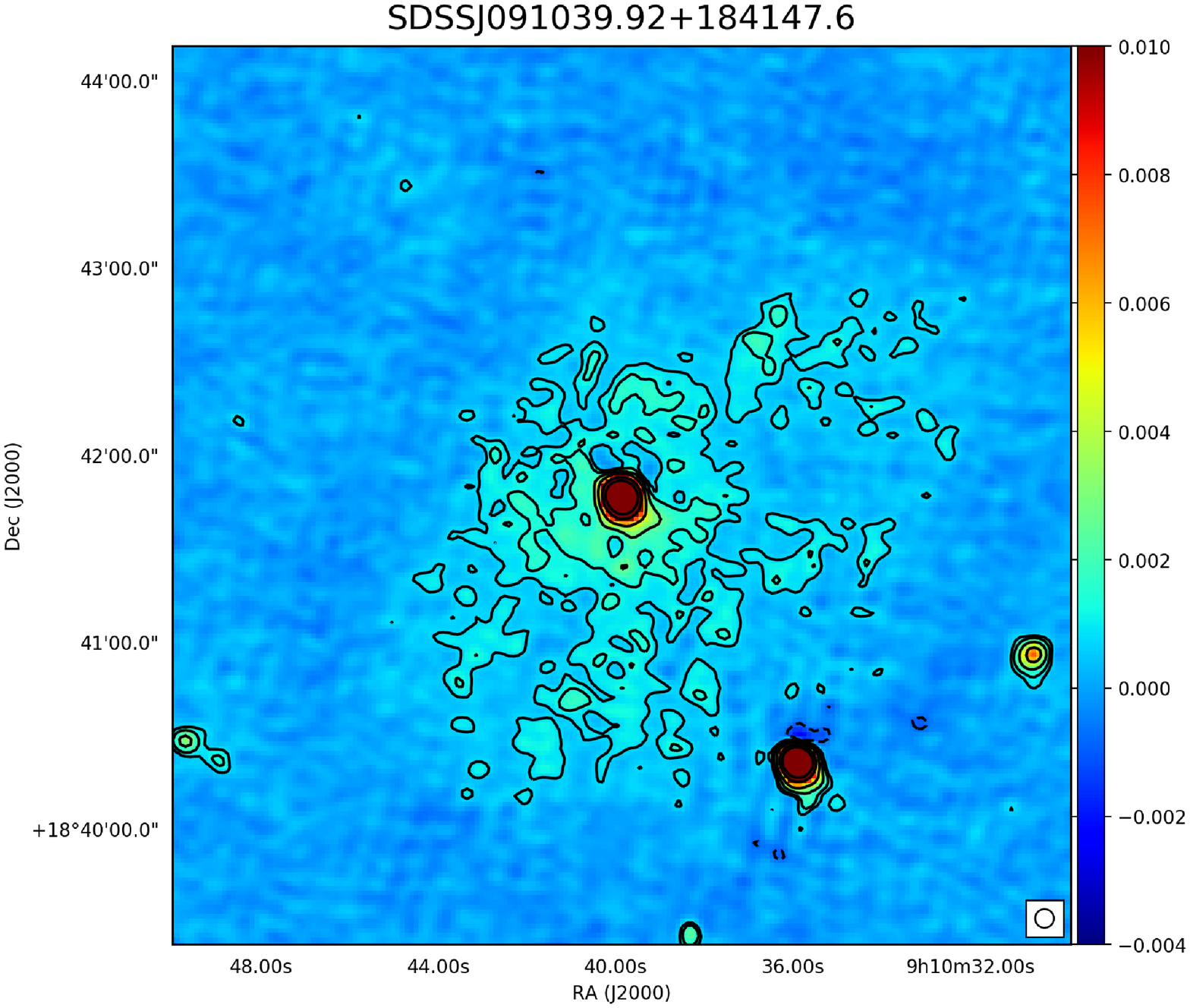}
\includegraphics[scale=0.38,angle=0]{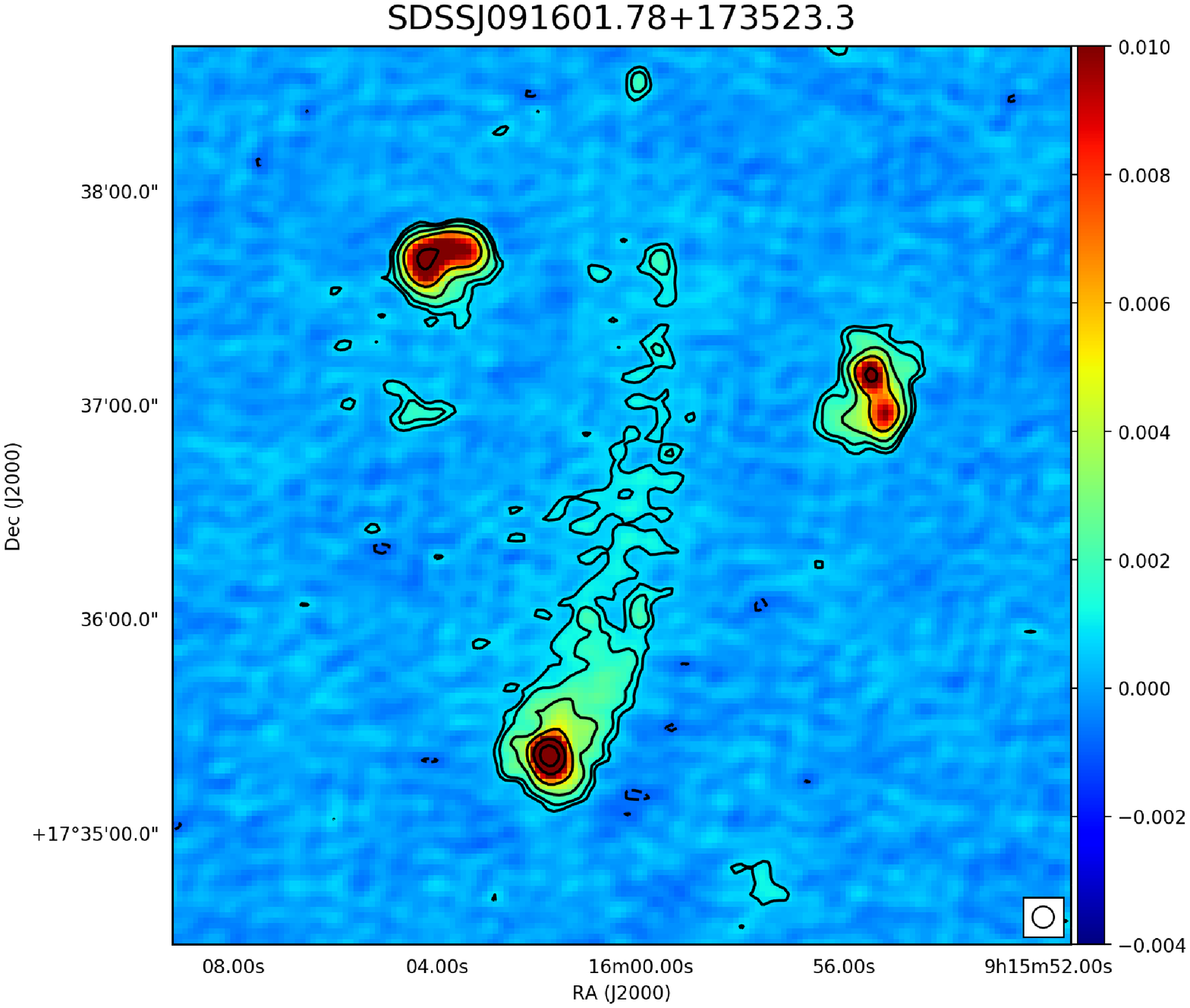}
\includegraphics[scale=0.38,angle=0]{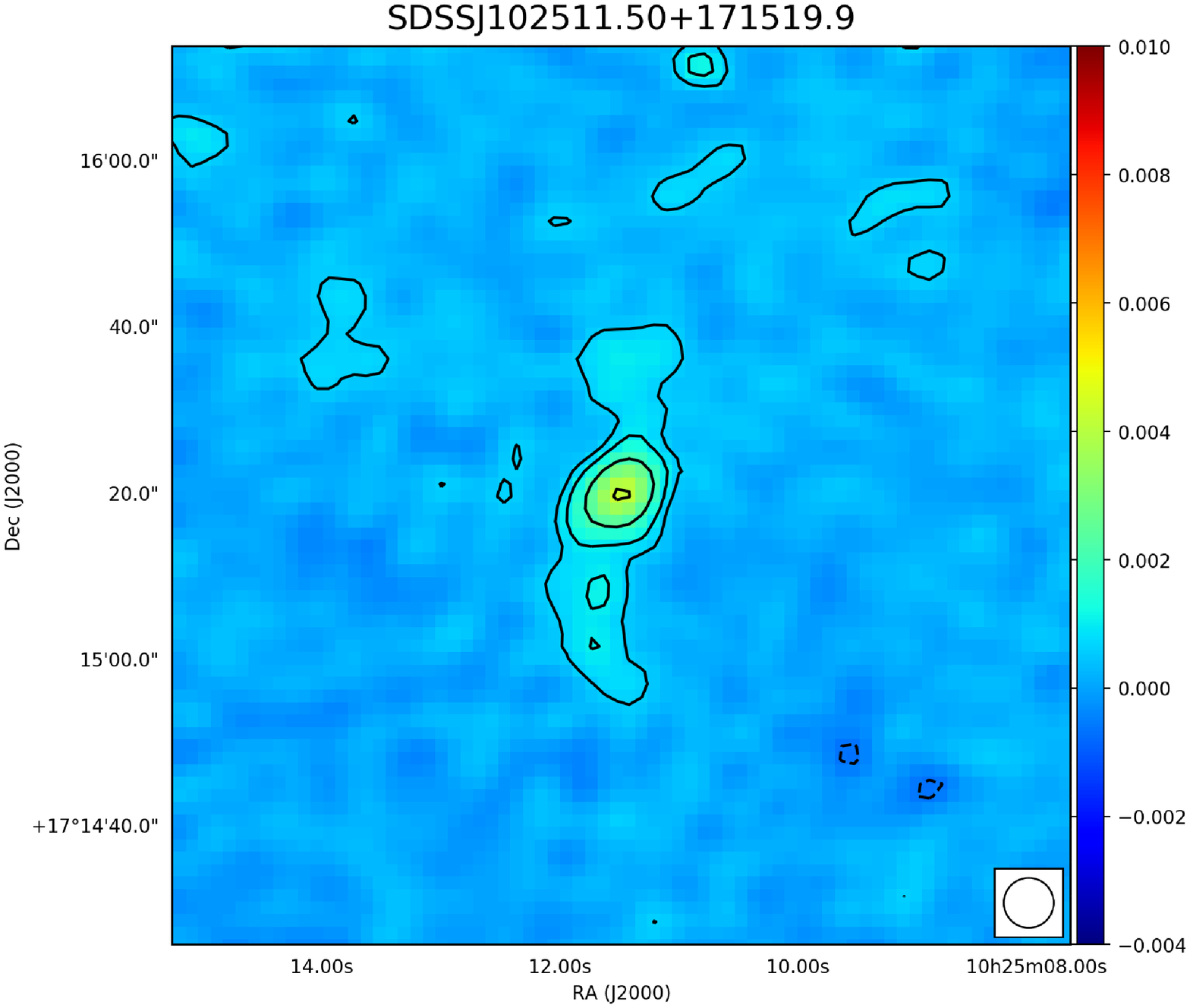}
\includegraphics[scale=0.38,angle=0]{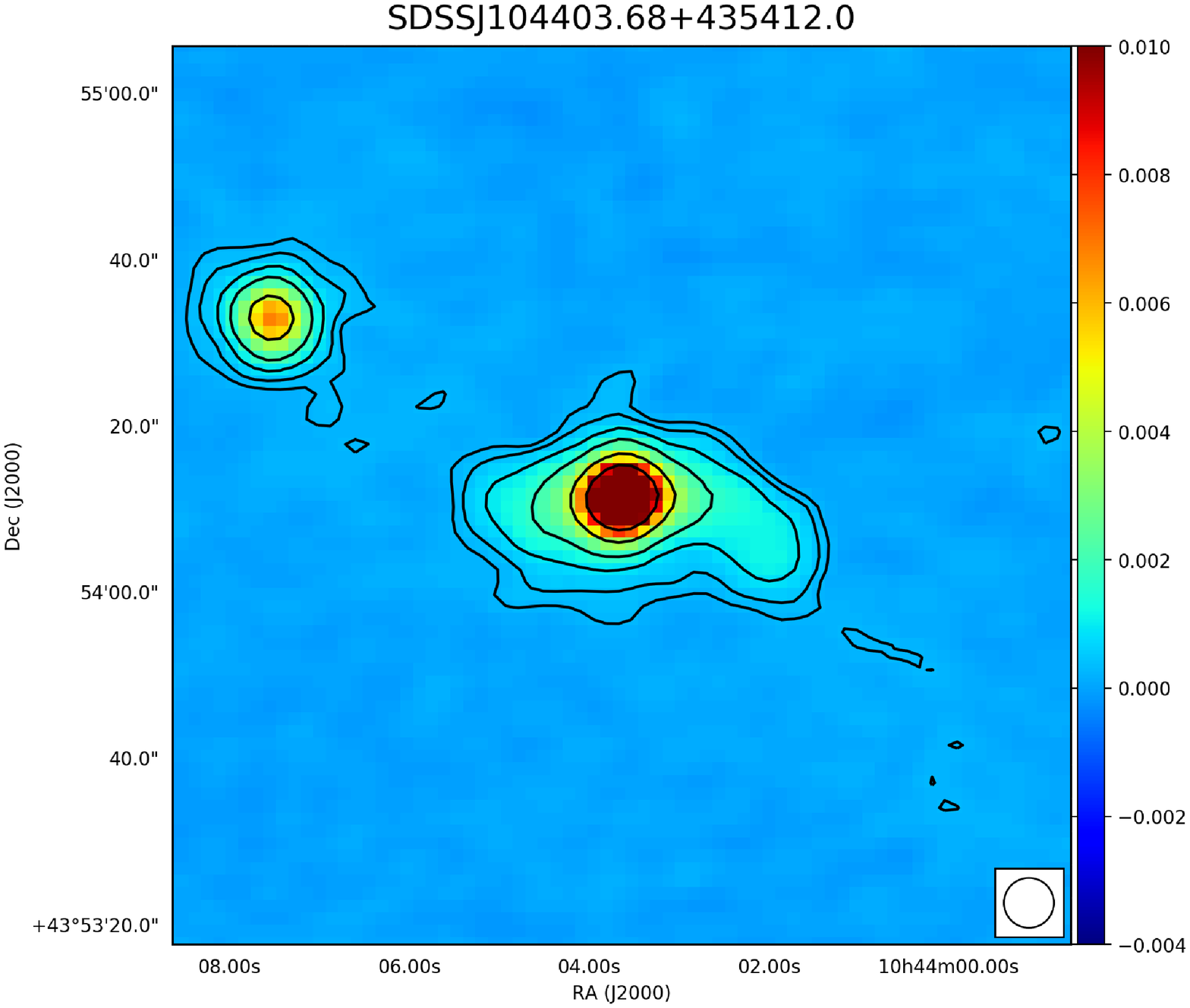}
\includegraphics[scale=0.38,angle=0]{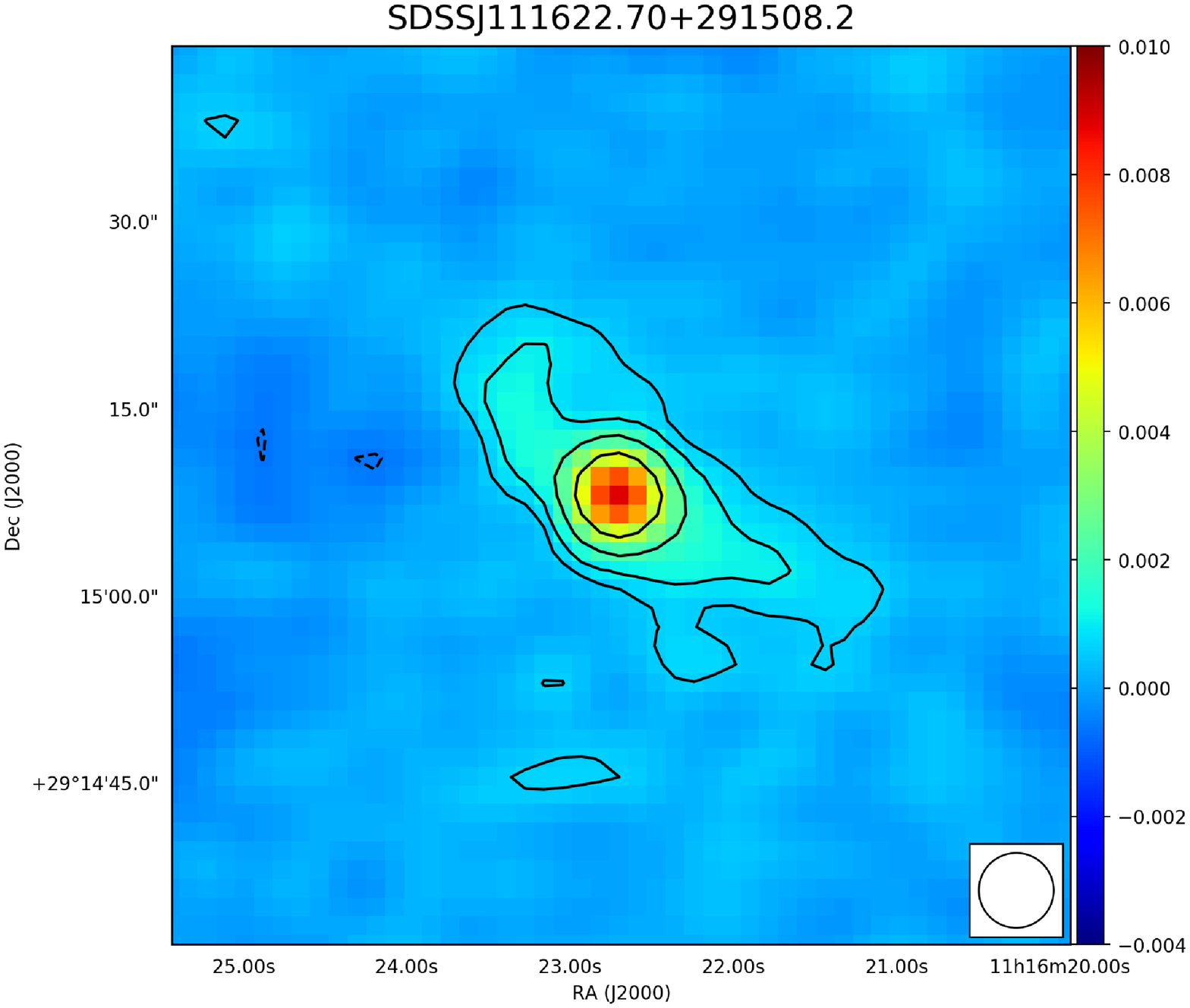}
 \caption{LOFAR images of the extended \FRo\ sources at a
    resolution of $\sim 6\arcsec$. The contour levels
    follow the sequence -3, 3, 5, 10, 20, 50, 100 $\sigma$, where
    $\sigma$ is the local r.m.s., as reported in Table \ref{tab},
    0.20, 0.20, 0.27, 0.21, 0.09, and 0.20 mJy beam$^{-1}$,
    respectively.}
\label{estese1} \end{figure*}

\begin{figure*}
\includegraphics[scale=0.38,angle=0]{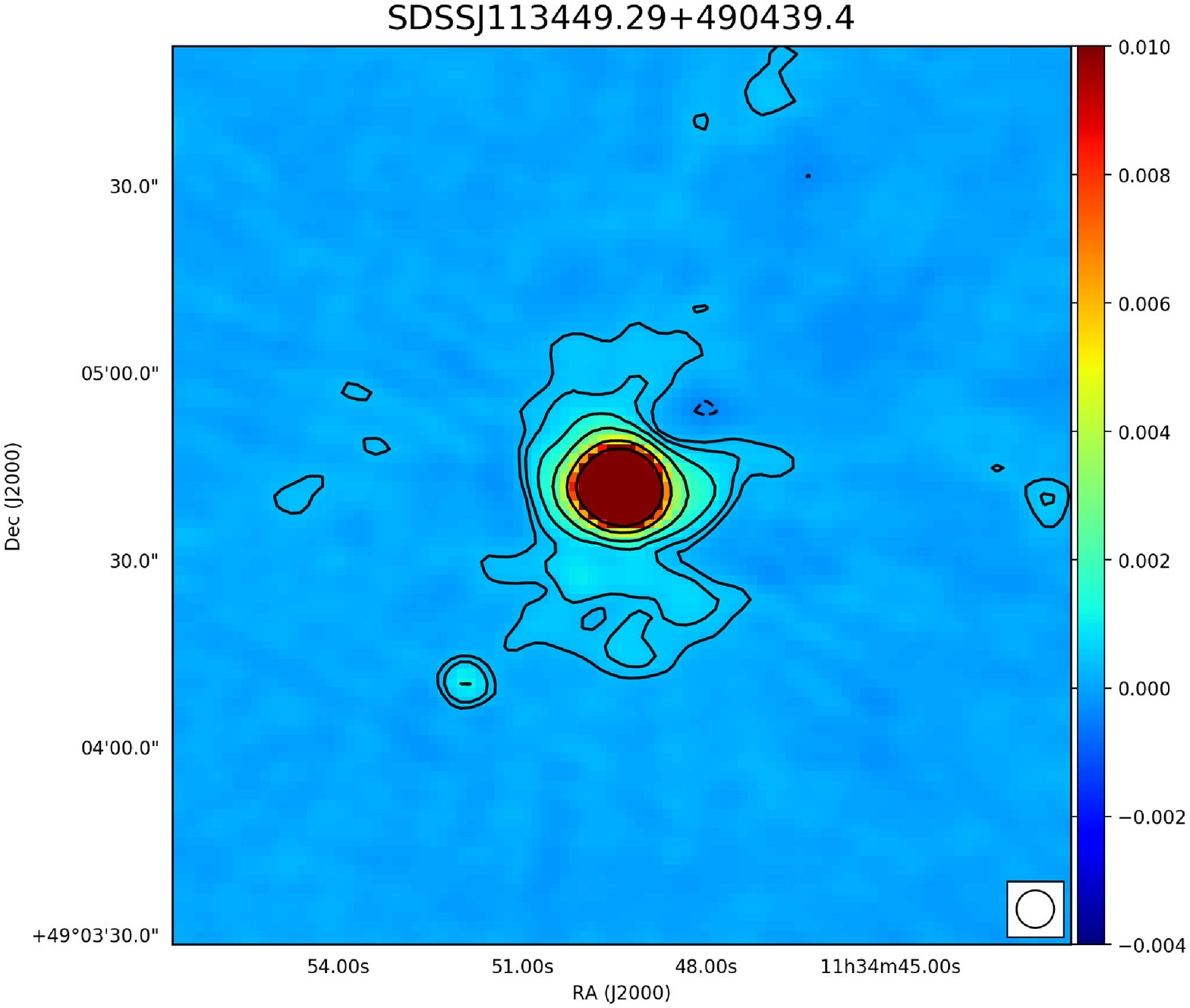}
\includegraphics[scale=0.38,angle=0]{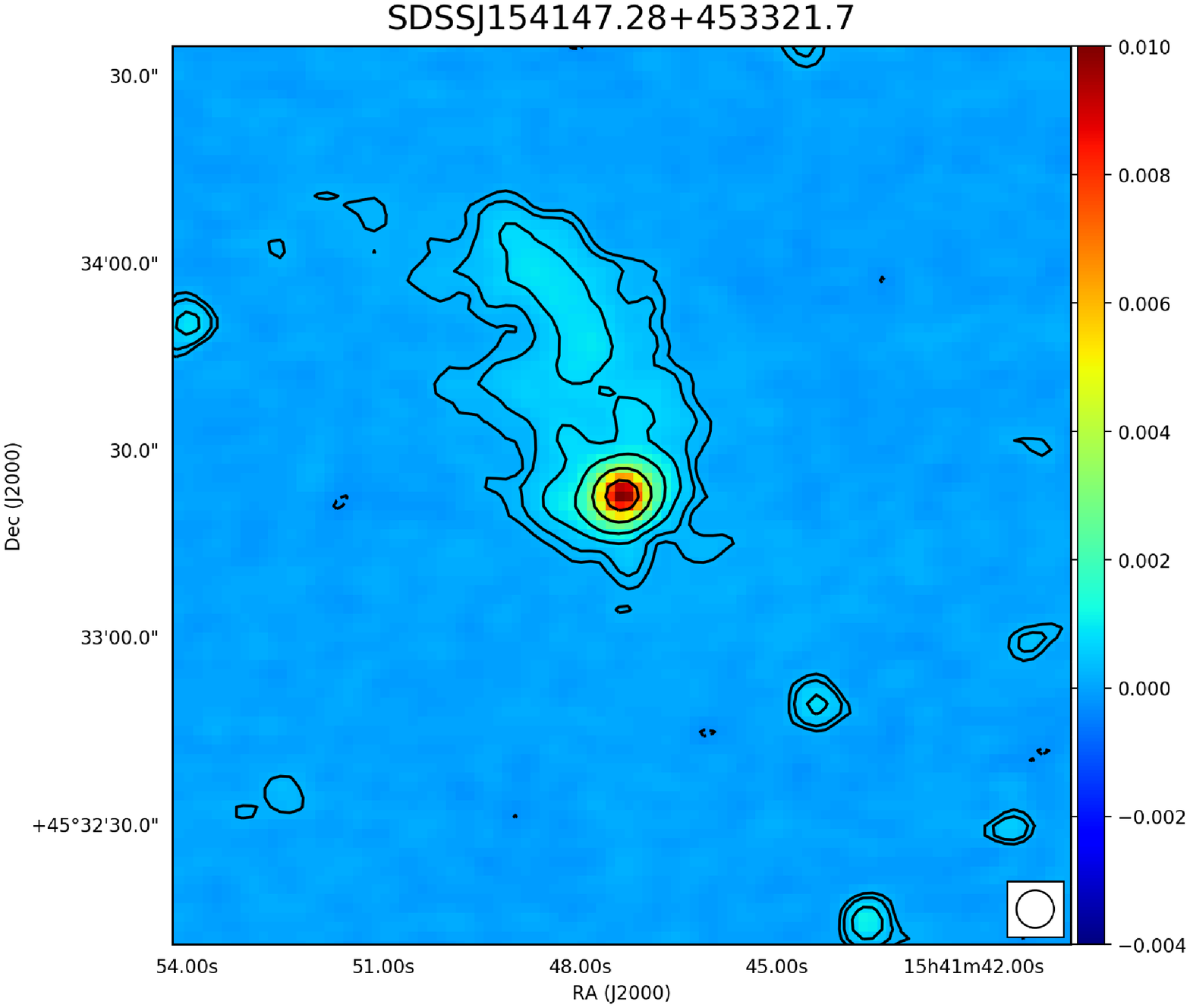}
\includegraphics[scale=0.38,angle=0]{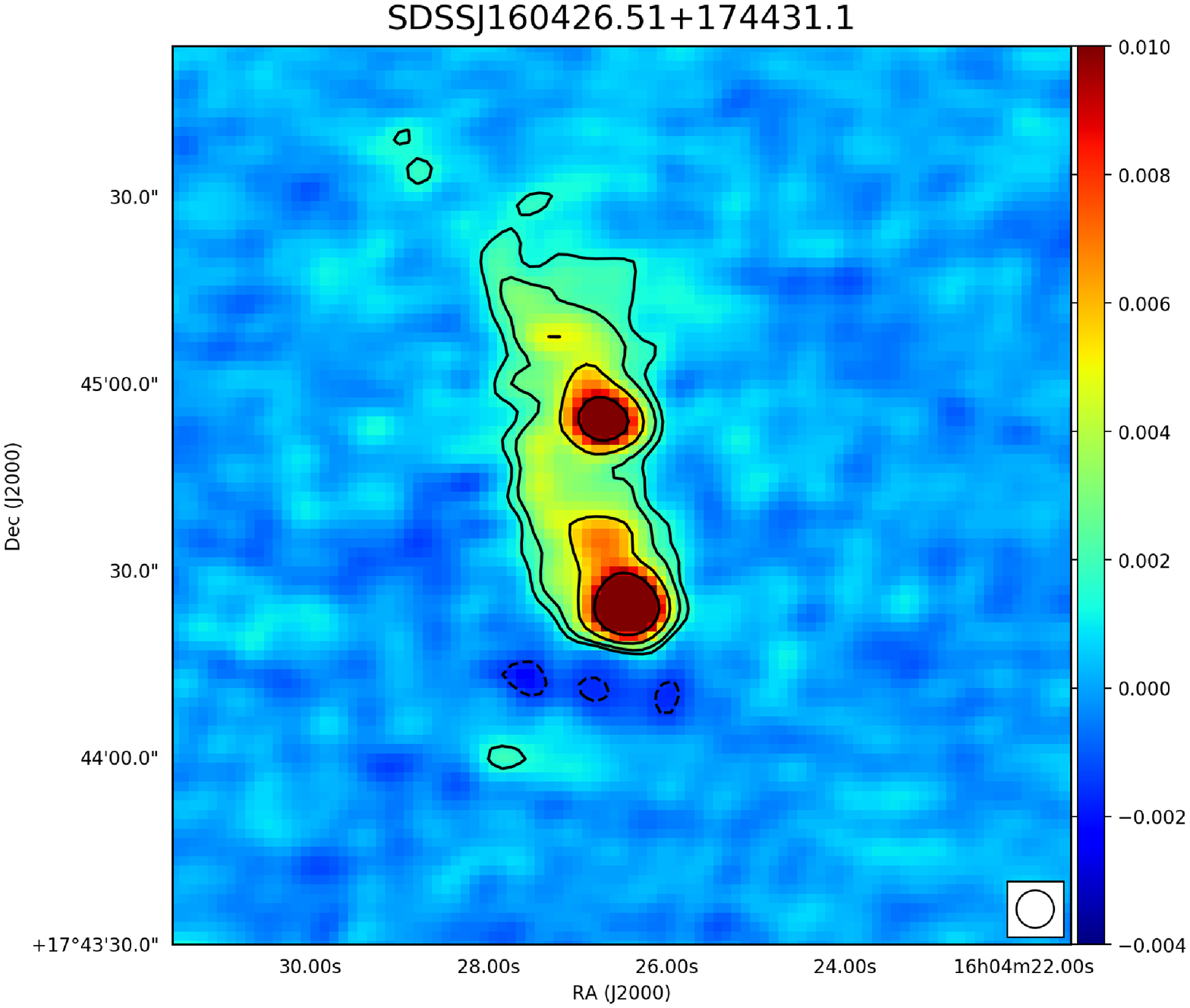}
\includegraphics[scale=0.38,angle=0]{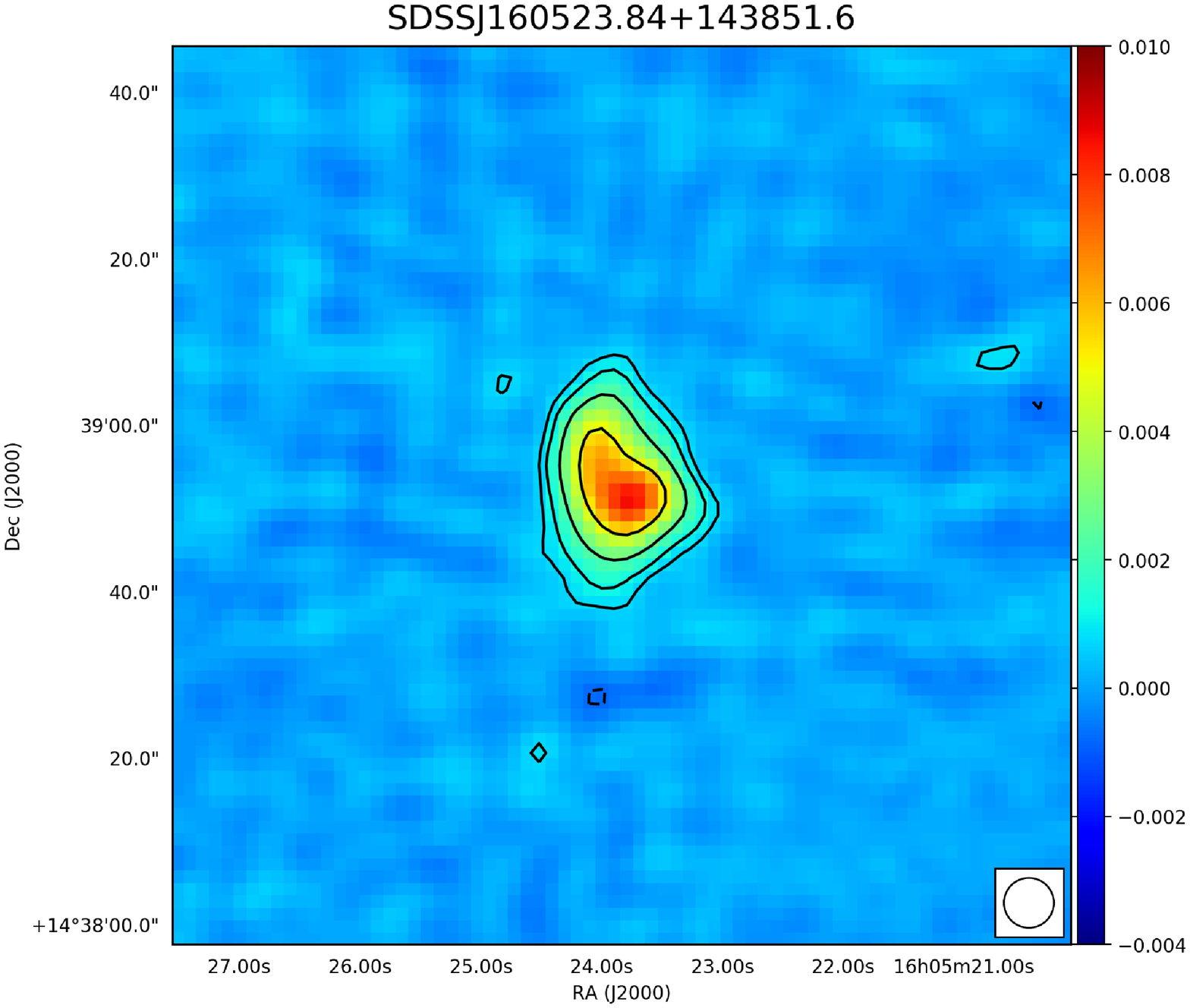}
\includegraphics[scale=0.38,angle=0]{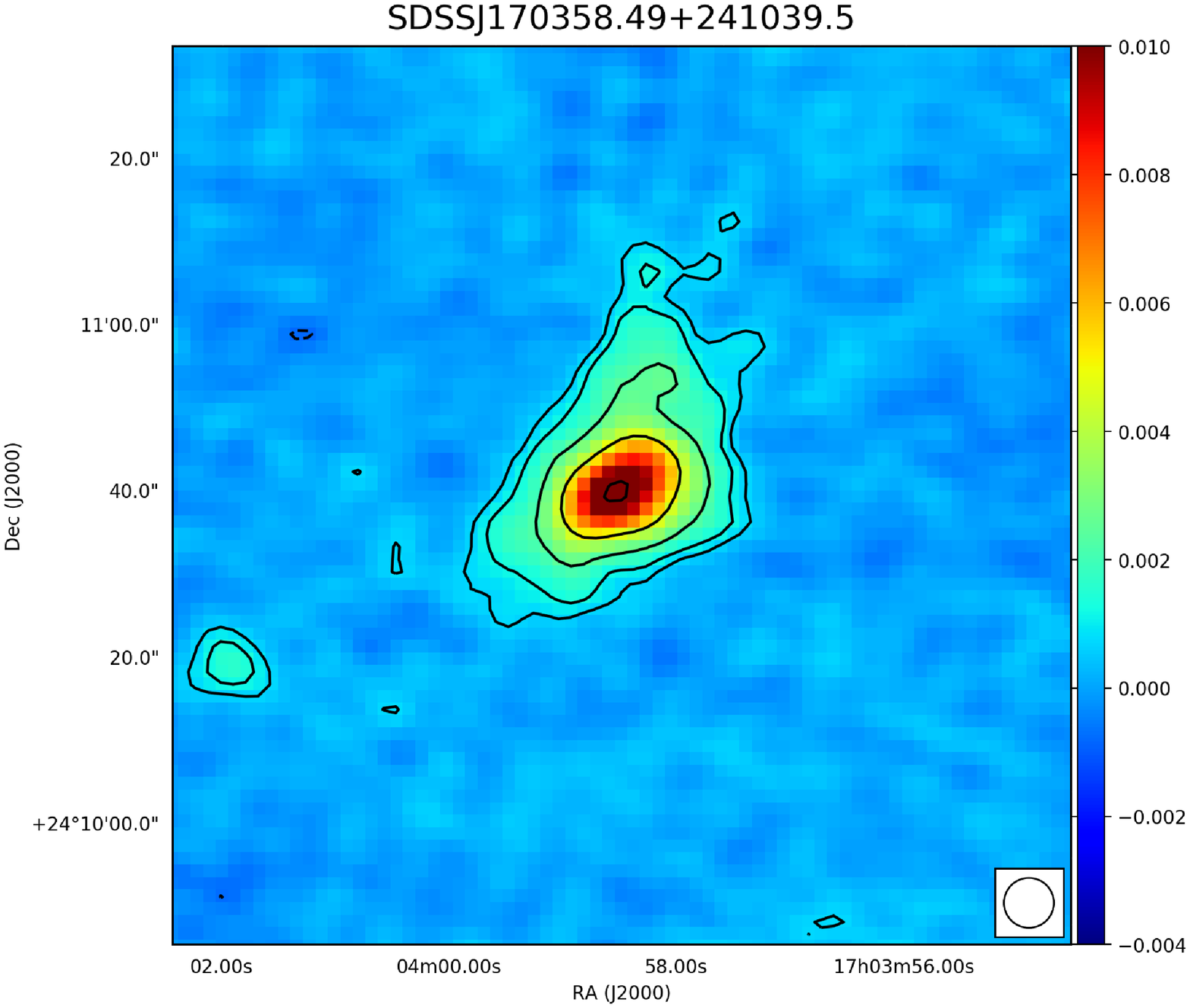}
\includegraphics[scale=0.38,angle=0]{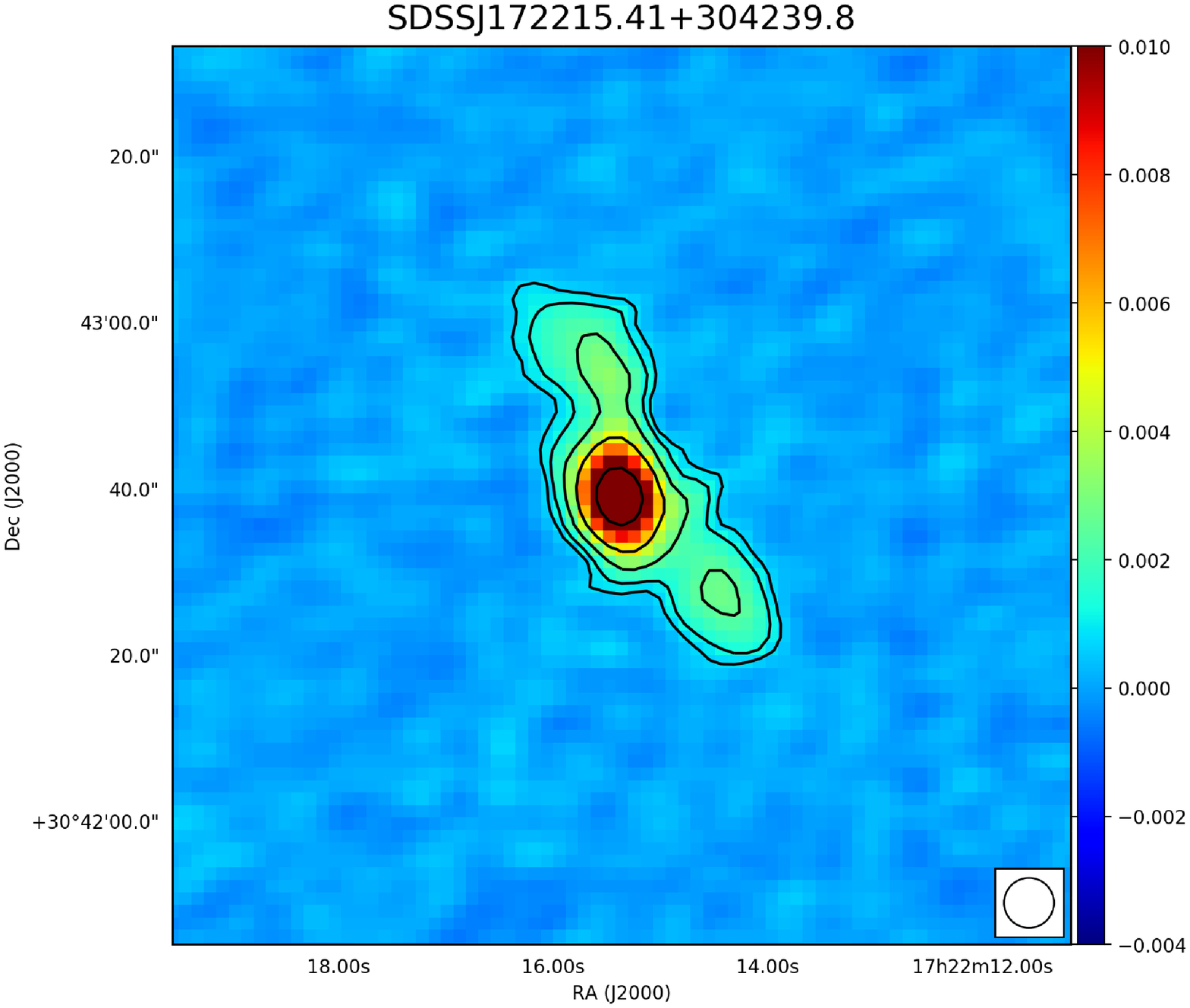}
  \caption{LOFAR images of the extended \FRo\ sources at a resolution
    of $\sim 6\arcsec$. The contour levels follow the
    sequence -3, 3, 5, 10, 20, 50, 100 $\sigma$, where $\sigma$ is the
    local r.m.s., as reported in Table \ref{tab}, 0.14, 0.10,
    0.38, 0.21, 0.22, and 0.10 mJy beam$^{-1}$, respectively.}
\label{estese2}
\end{figure*}

\begin{figure}
\includegraphics[scale=0.38,angle=0]{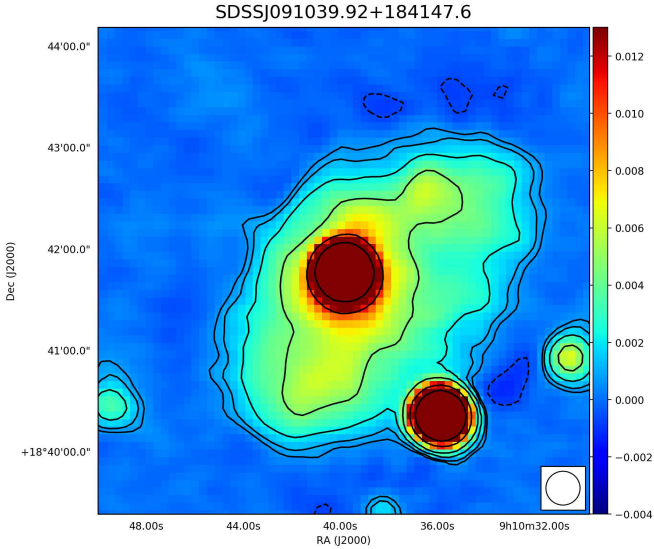}
\caption{LOFAR image at low resolution ($\sim 20\arcsec$) of J0910+18, best
  showing the elongated diffuse emission. The contour levels follow the
    sequence -3, 3, 5, 10, 20, 50, 100 $\sigma$, where $\sigma=500 \mu$Jy/beam.}
\label{altra2}
\end{figure}

\begin{figure}
\includegraphics[scale=0.38,angle=0]{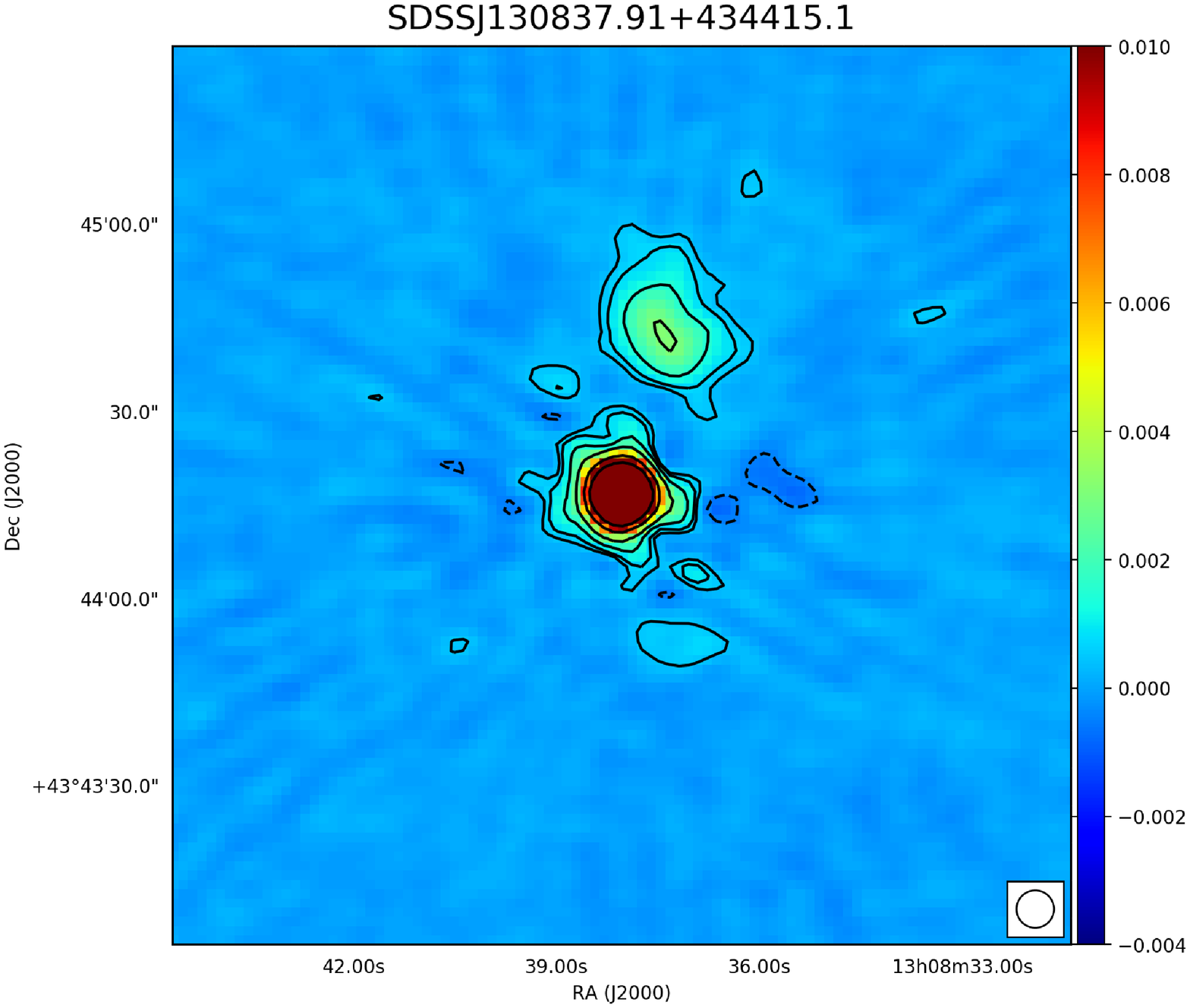}
\caption{LOFAR image of J1308+43 (resolution of $\sim 6\arcsec$)
  shows $\sim$ 30\arcsec\ north of the FR~0 core a detached diffuse
  radio structure. The contour levels follow the sequence -3, 3, 5,
  10, 20, 50, 100 $\sigma$, where $\sigma = 0.14$ mJy beam$^{-1}$.}
\label{altra1}
\end{figure}

\section{The LOFAR observations}
\label{lofar}

The LOFAR Two-metre Sky Survey (LoTSS, \citealt{shimwell17}) will
cover the whole northern sky with 3168 pointings of eight
hours\footnote{The low declination fields ($\delta \lesssim 20^{\circ}$)
  will instead be observed for 12 hours.} of dwell time each in the
frequency range between 120 and 168 MHz.
The LoTSS first data release (DR1) \citep{shimwell19} presented the
results obtained from observations of 424 square degrees in the HETDEX
Spring Field. The final release images were obtained by combining the
images from individual pointings of the survey, producing mosaics
covering the region of interest at a median sensitivity of 71
$\mu$Jy/beam. The flux density scale was adjusted to ensure
consistency with previous surveys (see \citealt{hardcastle16} for
further details).

The second LoTSS data release (DR2) will consist of two contiguous
fields at high Galactic latitude centered around 0h and 13h and
covering approximatively 5,700 square degrees (Shimwell et al. in
preparation). The DR2 provides fully calibrated mosaics at a
resolution of $\sim 6\arcsec$, catalogs, and it includes
42 FR~0s (marked as ``DR2'' in Tab. 1). FR~0s images can also be
obtained from individual LoTSS pointings, outside the DR2 area. We
restrict the analysis to the 24 objects (the ``P'' sources in Tab. 1)
with an offset from the field's center smaller than 3$^\circ$ in order
to limit ourselves to the regions of lower noise. The total number of
FR~0s with currently available LOFAR data is then 66, that is, about two
thirds of the FR0CAT sample.

We estimated the r.m.s. of each image in various regions, usually
centered 45$\arcmin$ from the FR~0s. For the FR~0s falling into the
DR2 the median r.m.s. is 85$\mu$Jy/beam, while this is 240$\mu$Jy/beam
for the ``P'' sources.

The photometry of the sources included in the DR2 is available from
the internally released catalog, while for the sources included in the
``P'' group we instead measured their flux density from their LOFAR
images. The flux density errors are dominated by the uncertainties in the
absolute calibration and are typically $\sim$10\%.

The list of FR~0s covered by LOFAR images is presented in Table. 1
where we indicate their SDSS name, their redshift, the local r.m.s. of
the image used, the flux density and luminosity at 150 MHz of the
central component from the LOFAR data (for the extended sources we
also give the total flux density and a morphological description), the
1.4 GHz flux density from FIRST, and the spectral index between these
frequencies. In the last column a code indicates
the origin of the image (``DR2'' or ``P'').

\section{Results}
\label{results}

\subsection{Morphology of FR~0s}

All \FRo\ sources with LOFAR observations are detected at 150 MHz.
Most of them are point-like with a deconvolved major axis $\lesssim 6
\arcsec$ (the median value is 5\farcs0), that is, 4 kpc at their median
redshift.

There are 12 clearly extended sources (see Table \ref{tab} and
Fig. \ref{estese1} and \ref{estese2}).\footnote{Somewhat surprisingly,
  the fraction of extended sources is lower (12\%) in the deeper DR2
  observations than in those covered by the shallower single survey's
  pointings (29\%).} The most commonly observed morphology among them
(in five sources, namely, J1025+17, J1044+43, J1116+29, J1134+49, and
J1722+30) is the presence of two rather symmetric jets, usually bent
with an S-shape structure, with a total extent between 20 and 40 kpc.
In three sources (J0916+17, J1541+45, and J1604+17) we see a head-tail
structure reaching $\sim$50 kpc; in one of them (J1604+17) a second
source is seen $\sim$30$\arcsec$ north, along the radio tail,
associated with a spiral galaxy interacting with the host of the radio
source. Two \FRo\ sources (J0807+14 and J1605+14) have instead a
core-jet shape, while one (J1703+24) is barely resolved and of
uncertain morphology.  The higher resolution observations of this
source at 1.4 GHz presented by \citet{baldi19} show a double-lobed
structure, with a total extent of $\sim$ 15 kpc. The last object
(J0910+18) is quite peculiar with a bright central source superposed
to a large scale plateau of diffuse radio emission. The low resolution
(20\arcsec) LOFAR image (see Fig. \ref{altra2}) reveals that this
plateau is just the central part of an elongated radio structure,
extending over $\sim 4^\prime$, that is, $\sim$150 kpc. In addition,
J1308+43 (see Fig. \ref{altra1}) has a low brightness diffuse feature
centered 30$\arcsec$ ($\sim$ 18 kpc) to the NNW side, but in this case
it is detached from the radio core and not obviously associated with
the FR~0.

For the extended sources we measured both the flux density of the
central component and the total flux density, measured integrating in
the region included by the 3$\sigma$ isophote. The central component
always accounts for a large fraction of the total emission, from 40\%
to 93\%. The luminosities of the extended emission range from
$10^{22}$ to $3\times10^{23} \WHz$. These structures were not visible
in the TGSS images.

\begin{figure}
\includegraphics[scale=0.55]{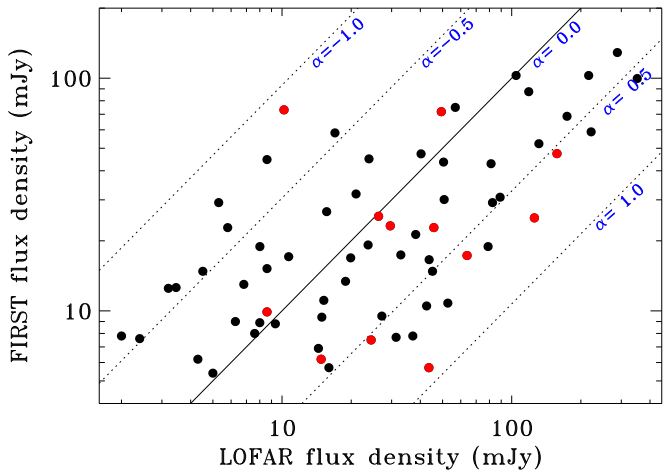}
\caption{Comparison of the flux densities of the compact components of
  the FR0CAT sources in LOFAR (at 150 MHz) and FIRST (at 1.4 GHz). The
  red dots represent the 12 sources extended in the LOFAR images. The
  lines are the loci of constant spectral indices $\alpha$ (defined as
  $F_{\nu}\propto\,\nu^{-\alpha}$) at the values indicated.}
\label{first}
\end{figure}

\begin{figure}
\includegraphics[scale=0.55]{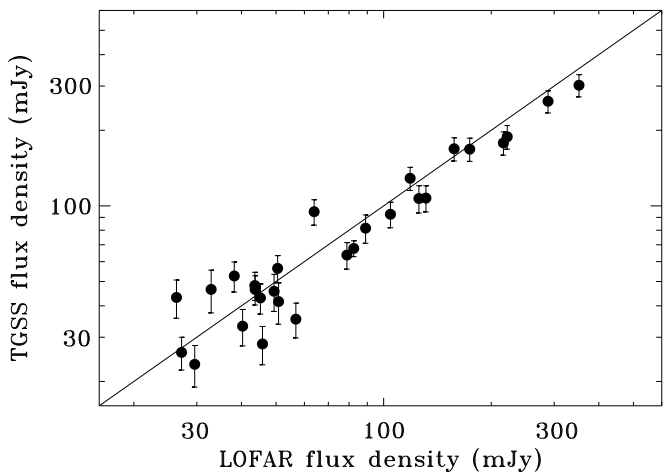}
\caption{Comparison of the flux densities of the 28 \FRo\ detected by
  TGSS and observed by LOFAR. The errors on the LOFAR fluxes, not
  reported, are dominated by the $\sim$10\% uncertainty in the
  absolute calibration.}
\label{tgss}
\end{figure}

None of the extended structures detected in the LOFAR images of the
\FRo\ sources has a counterpart in the FIRST images.  This is not
surprising if they are steep spectra features ($\alpha = 0.7$) given
their brightness levels at 150 MHz (0.6 to 1.2 mJy beam$^{-1}$)
leading to an expected surface brightness of 0.1 - 0.2 mJy beam$^{-1}$
at 1.4 GHz. The lower limits to the spectral indices of the extended
emission do provide strong constraints, with typical limits $\alpha
\gtrsim 0.4 - 0.6$, except for the head-tail source J0916+17 ($\alpha
\gtrsim 0.9$). However, although the LOFAR spatial resolution ($\sim
5\arcsec$) is very similar to that achieved by FIRST, in the latter
survey the structures larger than $\sim 1\arcmin$ might be at least
partly resolved out, given the lack of short baselines in the $u-v$
coverage \citep{helfand15}. An indication that this is indeed the case
comes from the ratio between the NVSS and FIRST flux densities of the
12 FR~0s extended at 150 MHz that, in seven\footnote{Namely, J0916+17,
  J1044+43, J1134+49, J1541+45, J1604+17, J1703+24, and J1722+30.} of
them, is $ 1.30 < F_{\rm NVSS}/F_{\rm FIRST} < 1.45 $: similarly high
ratios are measured only in 14/104 FR~0s of the complete sample
\citep{capetti17}. However, variability, combined with a high
  core dominance, prevents us from using the NVSS/FIRST flux ratio to
  extract robust measurements of the spectral slope of the extended
  emission. In fact, FR~0s show signs of variability over a timescales
  of a few years. This conclusion is based on multi-epoch VLBI images
  \citep{cheng18} and on the presence of a large number of FR~0s in
  which the flux density in the FIRST images significantly exceeds
  that measured from the NVSS \citep{baldi18}. Furthermore, hints of
  variability are also identified when comparing the 150-MHz LOFAR
  flux densities with the TGSS flux densities at the same frequency,
  as shown in Fig. \ref{tgss}.

We estimate the limit on the flux density of any extended emission at
150 MHz around the remaining FR~0s by considering an area of 100 kpc
$\times$ 100 kpc, the typical size of the edge-darkened FR~I sources
(forming the sample named \FR, \citealt{capetti17}) selected from the
same catalog of RGs from which we extracted the FR~0s. At the median
redshift of the \FRo\ of 0.037, this corresponds to $\sim 135\arcsec
\times 135\arcsec$. The median r.m.s. is 0.15 mJy beam$^{-1}$ and this
leads to an upper limit over this area of $\sim 10$ mJy, corresponding
to a luminosity of $\lesssim 5 \times 10^{22}$ W Hz$^{-1}$.

\subsection{Radio spectral properties of FR~0s}

In Fig. \ref{first} we compare the flux densities of the FR0CAT
sources in FIRST and LOFAR. The \FRo\ sources show a rather large
spread in spectral indices, ranging from $\alpha \sim -0.7$ to $\sim
0.8$, with a median value of $\overline\alpha \sim 0.1$. The fraction
of sources with a steep spectrum ($\alpha > 0.5$) is $\sim 20\%$, a
larger value with respect to what we obtained combining TGSS and NVSS
data ($\sim 8\%$, \citealt{capetti19}). The comparison of the spectral
indices for the 29 sources observed by LOFAR and detected by TGSS
indicate that in the majority of them (25) the difference in spectral
slope is smaller than 0.1, with no systematic offset. In the remaining
four sources the difference is larger, reaching
$\Delta\alpha=0.4$. This effect is likely due to the flux losses in
the FIRST and TGSS images, with respect to the NVSS and LOFAR
observations, respectively.

The central components associated with the 12 extended sources are all
among the upper half of the luminosity distribution of FR~0s, having a
median luminosity of $10^{23} \WHz$, and a median slope of
$\overline\alpha \sim 0.4$.

\section{Discussion}

The definition of compact radio sources, at the base of our study of
FR~0s, clearly depends on the spatial resolution, depth, and frequency
of the available images. \citet{baldi19} presented high resolution
images (between 0\farcs3 and 1\arcsec) of a subsample of 18
\FRo\ sources, detecting extended emission (on a scale between 2 and
14 kpc) only in four of them. The LOFAR observations provide us with
additional evidence that the inclusion of a radio source into the
FR~0s class is nearly independent of the characteristics of the data
used and that, consequently, this class of radio galaxies is rather
well defined. In particular, they confirm that FR~0s are in general
compact sources, of high core dominance. Even when extended radio
emission is detected in the FR~0s, thanks to the deep LOFAR images, it
represents a small fraction of the total flux densities in these
objects, ranging from $\sim 5$ to $\sim 40\%$. For the unresolved
FR~0s, we set a limit to the luminosity of any extended emission of
$\lesssim 5 \times 10^{22}$ W Hz$^{-1}$ within an area of 100 kpc
$\times$ 100 kpc. As reference, the \FR\ sources in \citet{capetti17}
have predicted luminosities at 150 MHz (by assuming a spectral slope
of 0.7 between 150 and 1400 MHz) in the range $\sim 10^{24} - 10^{26}$
W Hz$^{-1}$.

The properties of the 12 extended sources in the LOFAR images can be
compared with those of the FR~Is selected from the FIRST images
\citep{capetti17}. Their median optical magnitude is $M_r = -22.38$, a
value intermediate between that measured for the \FRo\ and
\FR\ samples ($M_r = -22.05$ and $M_r = -22.52$, respectively). From
the point of view of the environment, the median number of
cosmological neighbors\footnote{``Cosmological neighbors'' are defined
  as the galaxies lying within a projected radius of 2 Mpc and having
  a spectroscopic redshift $z$ differing by less than 0.005 from the
  radio galaxy in the center of the field examined.} is $N_{\rm
  cn}^{2000}$= 25 for the extended sources, to be compared to 13 and
44 for the \FRo\ and \FR\ objects, respectively. Conversely, these 12
FR~0s do not differ from the remaining sources from the point of view
of optical line properties. This analysis indicates that large scale
radio structures are preferentially seen in FR~0s whose host galaxies
are more luminous and are located in a denser environment than
average. As already mentioned in the Introduction, the radio structure
of most FR~0s is confined within the core of their hosts and is not
directly affected by the larger scale environment. Nonetheless, the
larger structures that a minority of FR~0s are able to form might be
easier to detect in a denser environment where their expansion and the
adiabatic losses are reduced. The role of host's mass and environment
in determining the properties of a radio source is still unclear, but
these results strengthen the link between these two parameters with
the jet power and the ability of the AGN to produce large scale radio
emission.

Almost all of the \FRo\ sources with extended radio emission (detected
with either LOFAR or with the VLA) show the typical morphology usually
associated with low-power jets: collimated structures, albeit often
distorted, of rapidly declining brightness at increasing distance from
the nucleus. This result argues against the possibility that these
structures are relic emission from previous phases of activity. In
such objects the radio emission is generally characterized by a
diffuse morphology (see \citealt{brienza16}), due to the lateral
expansion that occurs when the jets cease to replenish the large scale
structures.

The presence of jets in FR~0s can also be argued, besides their direct
morphological detection, from their spectral radio properties. The
fraction of sources with a steep spectrum ($\alpha > 0.5$) is $\sim
20\%$, represents an indication of the presence of substantial
extended emission confined within the LOFAR spatial resolution limit
($\sim$3-6 kpc). This suggestion is confirmed by considering the
sources showing extended emission in the high resolution VLA data of
\citet{baldi19}. Three of these four sources have LOFAR observations:
J1703+24 is extended in both datasets, while J1213+50 and J1559+44 are
both point-like in the LOFAR images. While the first source has a size
of $\sim12\arcsec$ (9 kpc) at 1.5 GHz, the last two objects only
extend for $\sim2\arcsec$, well below the LOFAR resolution. All of
them are relatively high power sources (between 0.3 and 1.0$\times
10^{23} \WHz$ at 150 MHz) and with a spectral index $\alpha \sim
0.3$. Besides the four FR~0s with jetted morphology, the VLA images of
\citeauthor{baldi19} indicate that there are three further objects
(out of 18) with a steep ($\alpha > 0.5$) spectrum between 1.4 and 4.5
GHz. These are unresolved objects, with sizes $\lesssim 0\farcs3$
($\lesssim$ 0.2 kpc), with a dominant contribution from the extended
emission from their small scale jets. 

The overall fraction of FR~0s showing evidence for the presence of
jets can be estimated by including in the census those in which they
are directly detected based on the available LOFAR, VLA, or VLBI
\citep{cheng18} images and those in which the spectral slope indicates
a dominance of an optically thin component, that is, those with $\alpha >
0.5$. By considering the overlap between these classes we obtain a
lower limit to the fraction of jetted FR~0 of $\gtrsim 40\%$.

The morphologies of the extended structures observed with LOFAR in the
FR~0s recall what has been recently observed in other radio galaxies
and in particular in the radio source associated with NGC~3998
\citep{sridhar20}. This galaxy produces a low-power and highly core
dominated source with two elongated and distorted lobes of low-surface
brightness with a total size of $\sim 30$ kpc. The overall properties
of NGC~3998 are consistent with a FR~0 classification, very similar to
the \FRo\ sources with extended structures we found in this study. The
spectral slope between 150 MHz and 1.4 GHz of the extended emission in
NGC~3998 is $\alpha \sim 0.6$ suggesting that it is still actively fed
by the AGN and it is not a relic structure. It would be clearly very
important to measure the spectral indices of the extended emission in
the FR~0s to test our suggestion that they are sources actively fed by
their jets, but the depth of the available higher frequency images is
not sufficient to provide strong constraints.

Other classes of AGN show radio structures reminiscent of those seen
in FR~0s. In particular it has long been known that Seyfert galaxies,
despite their definition as radio-quiet objects, often show extended
radio emission (for example, \citealt{ulvestad84,nagar02,nagar05}) extending
over a few kpc and with a high core dominance. More recently,
\citet{baldi18lem} found that most nearby LINERs are radio emitters,
albeit of very low-power (typically $\sim 10^{20} \WHz$). They are
often associated with radio-jets and are also highly
core-dominated. LINERs share with the FR~0s the typical range of
values of optical line ratios, used for the spectroscopic
classification (see, \citealt{buttiglione10}).  FR~0s, FR~Is, and
LINERs also have a similar ratio between the core and the optical line
luminosity, an indication of a similar efficiency of the central
engine to produce highly relativistic electrons.  These similarities
are particularly interesting considering that the hosts of these
radio-quiet AGN include late-type galaxies and low-mass ellipticals.

\section{Summary and conclusions}
\label{summary}

We explored the low-frequency radio properties of the sample of
compact radio sources associated with nearby ($z<0.05$) massive
early-type galaxies, FR0{\sl{CAT}}, by using LOFAR observations at 150
MHz, available for 66 out of 104 FR~0s: all of them are detected,
usually showing point-like structures. Resolved radio emission of low
surface brightness is detected in 12 FR~0s: it contributes from
between 5\% and 40\% of the total radio power at 150 MHz. The LOFAR
observations confirm the general paucity of large scale emission in
FR~0s, as already indicated by the FIRST images used for the selection
of the sample. The extended radio emission usually have a jetted
morphology extending between 15 and 40 kpc. In the remaining FR~0s we
set an upper limit to any extended emission of $\lesssim 5 \times
10^{22}$ W Hz$^{-1}$, a factor 10 - 10$^3$ below the typical
luminosity of FR~Is. It is likely that the FR~0s in which we detected
large scale emission are just the tip of the iceberg of a broad
distribution of extended power.

The spectral slopes of FR~0s between 150 MHz and 1.4 GHz span a broad
range ($-0.7 \lesssim \alpha \lesssim 0.8$, median $\overline\alpha
\sim 0.1$); 20\% of them have steep spectra ($\alpha \gtrsim 0.5$), an
indication of the presence of substantial extended emission confined
within the spatial resolution limit ($\sim$3-6 kpc at
$z\sim0.05$). The fraction of FR~0s showing evidence for the presence
of jets, by including both spectral and morphological information, is
at least $\sim 40\%$.

Our study confirms that FR~0s and FR~I can be interpreted as two
extremes of a continuous population of radio sources characterized by
a broad distribution of sizes and luminosities of their extended radio
emission, from low-luminosity compact RGs to Mpc-scale FRI~s (see,
\citealt{mingo19}). In this context, the widespread presence of
jets in FR~0s, an indication derived in a substantial fraction of the
\FRo\ sample, either from the morphology of their resolved structures
or from their spectral shape, is of great importance. FR~0s thus
represent the low-end in size and power of jetted radio sources. The
properties of these sources confirm the continuity of the properties
of low-power radio galaxies starting from the compact FR~0s,
characterized by a paucity of extended emission, small linear sizes,
and high values of core dominance, to the most powerful FR~Is which
can extend for hundreds of kpc. Most likely, these differences are
driven by the different properties of their jets, being lighter and/or
slower in FR~0s, leading to a lower momentum flux. For this reason
they are more subject to the effects of instabilities, turbulence, and
entrainment causing their premature disruption and limiting their
expansion to subgalactic scales.

Alternatively, FR0s might experience recurrent short-phases of
activity. Large scale radio structures are preferentially seen in
FR~0s whose host galaxies are more luminous and are located in a
denser environment than average. The role of host’s mass and
environment in determining the properties of a radio source are
unclear, but these results suggest the presence of a connection
between these two parameters with the jet power. For example, short
activity phases might be more commonly triggered in a poor group
environment due to the smaller amount hot gas available in comparison
with cluster of galaxies.

Various questions about the FR~0s remain open and require further
studies. The fraction of young radio galaxies among the FR~0s is not
well constrained. Such objects must exist as they represent the early
phase of the evolution of the extended radio sources and carry
important information about the onset of nuclear activity. Young radio
sources (including, for example, compact steep and giga-hertz peaked sources)
are characterized by high curvature convex spectra, with a low
frequency turn-over due to absorption. While we have now adequate low
frequency data, the overall spectral shape of FR~0s can not be
properly studied due to the lack of sufficient information at higher
frequencies. Similarly, higher resolution observations are required to
detect and isolate the jet emission and to compare their properties
(measuring the jets asymmetry, related to their speed) with
those of the FR~Is. The LOFAR international baselines will play a key
role in this research, because at 150 MHz the contrast between the
steep jets and the flat cores is enhanced with respect to higher
frequencies. Finally, deep observations at a resolution similar to
that obtained with LOFAR are needed for a measurement of the spectral
indices of the extended emission of FR~0s and to discriminate between
sources actively fed by their jets and relic emission.

\begin{acknowledgements}
MB acknowledges support from the ERC-Stg DRANOEL, no 714245. MJH
acknowledges support from the UK Science and Technology Facilities
Council (ST/R000905/1). HR acknowledges support from the ERC Advanced
Investigator programme NewClusters 321271. PNB is grateful for support
from the UK STFC via grant ST/R000972/1.

This paper is based on data obtained from the International LOFAR
Telescope (ILT). LOFAR \citep{vanhaarlem13} is the Low-Frequency Array
designed and constructed by ASTRON. It has observing, data processing,
and data storage facilities in several countries, which are owned by
various parties (each with their own funding sources), and are
collectively operated by the ILT foundation under a joint scientific
policy. The ILT resources have benefitted from the following recent
major funding sources: CNRS-INSU, Observatoire de Paris and
Universit\'{e} d'Orl\'{e}ans, France; BMBF, MIWF-NRW, MPG, Germany;
Science Foundation Ireland (SFI), Department of Business, Enterprise
and Innovation (DBEI), Ireland; NWO, The Netherlands; The Science and
Technology Facilities Council, UK; Ministry of Science and Higher
Education, Poland; Istituto Nazionale di Astrofisica (INAF),
Italy. This research made use of the Dutch national e-infrastructure
with support of the SURF Cooperative (e-infra 180169) and the LOFAR
e-infra group. The J\"{u}lich LOFAR Long Term Archive and the German
LOFAR network are both coordinated and operated by the J\"{u}lich
Supercomputing Centre (JSC), and computing resources on the
Supercomputer JUWELS at JSC were provided by the Gauss Centre for
Supercomputing e.V. (grant CHTB00) through the John von Neumann
Institute for Computing (NIC). This research made use of the
University of Hertfordshire high-performance computing facility and
the LOFAR-UK computing facility located at the University of
Hertfordshire and supported by STFC [ST/P000096/1], and of the LOFAR
IT computing infrastructure supported and operated by INAF, and by the
Physics Dept. of Turin University (under the agreement with Consorzio
Interuniversitario per la Fisica Spaziale) at the C3S Supercomputing
Centre, Italy.

\end{acknowledgements}

\bibliographystyle{./aa}

\begin{appendix}
\section{Radio properties of the sample}

\onecolumn
 \begin{longtable}{r r r r r r l r r r c }

\caption{Radio properties of the sample}
\label{tab} \\                                   

\hline 
SDSS~name & z & r.m.s. & \multicolumn{2}{c}{F(150)}  & Size & Morph. & L(150) & F(1.4)   & $\alpha$ & Im. type \\                                   
          &   &        & \multicolumn{2}{l}{Central Total}  &        &        &          &          &          \\                                   
\hline	
\endfirsthead

\multicolumn{3}{c}{{\tablename} \thetable{} -- Continued} \\                            [0.5ex]
\hline 
SDSS~name & z & r.m.s. & \multicolumn{2}{c}{F(150)}  & Size & Morph. & L(150) & F(1.4)   & $\alpha$ & Im. type \\                                   
          &   &        & \multicolumn{2}{l}{Central Total}  &        &        &          &          &          \\                                   
\hline
\endhead

\hline
  \multicolumn{10}{c}{{Continued on Next Page}} \\                                   
\endfoot

  \\                                   [-1.8ex] 

\endlastfoot

 J010852.48$-$003919.4 &0.045& 0.26 &  6.8  &      &  3.9$\pm$0.8 &               &22.52  & 13.0 & -0.29   &   P  \\ 
 J011204.61$-$001442.4 &0.044& 0.29 &  8.0  &      &  2.5$\pm$0.3 &               &22.57  & 18.9 & -0.38   &   P  \\ 
 J011515.78+001248.4   &0.045& 0.23 & 23.9  &      &  4.0$\pm$0.3 &               &23.06  & 45.0 & -0.28   &   P  \\ 
 J075354.98+130916.5   &0.048& 0.21 &  3.5  &      &  7.5$\pm$1.7 &               &22.28  & 12.6 & -0.58   &   P  \\ 
 J080716.58+145703.3   &0.029& 0.20 & 26.3  & 32.6 &  8.8$\pm$0.5 & Core-jet      &22.72  &  25.5&  0.01   &   P  \\ 
 J083158.49+562052.3   &0.045& 0.06 & 27.2  &      &  3.2$\pm$0.1 &               &23.12  &   9.5&  0.47   &  DR2 \\ 
 J084102.73+595610.5   &0.038& 0.06 & 15.2  &      &  4.1$\pm$0.1 &               &22.72  &  11.1&  0.14   &  DR2 \\ 
 J090652.79+412429.7   &0.027& 0.08 & 17.0  &      &  4.4$\pm$0.1 &               &22.47  &  58.2& -0.55   &  DR2 \\ 
 J090734.91+325722.9   &0.049& 0.09 & 81.3  &      & 10.2$\pm$0.1 &               &23.67  &  42.9&  0.29   &  DR2 \\ 
 J090937.44+192808.2   &0.028& 0.27 &174.1  &      &  3.9$\pm$0.2 &               &23.51  &  68.6&  0.42   &   P  \\ 
 J091039.92+184147.6   &0.028& 0.20 &157.4  &480.7 &  4.4$\pm$0.2 & Diffuse       &23.47  &  47.4&  0.54   &   P  \\ 
 J091601.78+173523.3   &0.029& 0.27 & 63.9  &157.6 &  5.9$\pm$0.6 & Head-tail     &23.11  &  17.3&  0.58   &   P  \\ 
 J093003.56+341325.3   &0.042& 0.09 & 89.0  &      &  5.0$\pm$0.1 &               &23.58  &  30.8&  0.48   &  DR2 \\ 
 J093938.62+385358.6   &0.046& 0.06 &  4.3  &      & 35.4$\pm$2.5 &               &22.34  &  6.2 & -0.16   &  DR2 \\ 
 J094319.15+361452.1   &0.022& 0.07 & 56.8  &      &  5.7$\pm$0.1 &               &22.82  & 74.9 & -0.12   &  DR2 \\ 
 J102403.28+420629.8   &0.044& 0.06 &  5.0  &      &  3.6$\pm$0.1 &               &22.37  &  5.4 & -0.03   &  DR2 \\ 
 J102511.50+171519.9   &0.045& 0.21 &  8.6  & 14.6 &  8.7$\pm$0.9 & Two-sided     &22.62  &   9.9& -0.06   &   P  \\ 
 J103719.33+433515.3   &0.025& 0.18 &288.4  &      &  4.8$\pm$0.1 &               &23.64  & 128.9&  0.36   &  DR2 \\ 
 J104403.68+435412.0   &0.025& 0.09 & 29.6  & 43.5 & 15.9$\pm$0.1 &Two-sided      &22.65  &  23.2&  0.11   &  DR2 \\  
 J104852.92+480314.8   &0.041& 0.10 & 78.9  &      &  5.0$\pm$0.1 &               &23.50  &  18.9&  0.64   &  DR2 \\ 
 J105731.16+405646.1   &0.025& 0.08 & 40.3  &      &  6.3$\pm$0.1 &               &22.78  &  47.3& -0.07   &  DR2 \\ 
 J111113.18+284147.0   &0.029& 0.10 & 50.5  &      &  3.1$\pm$0.1 &               &23.01  & 43.6 &  0.07   &  DR2 \\ 
 J111622.70+291508.2   &0.045& 0.20 & 10.2  & 16.5 &  4.7$\pm$0.3 & Two-sided     &22.69  &  73.1& -0.88   &  DR2 \\ 
 J111700.10+323550.9   &0.035& 0.08 & 23.7  &      &  5.0$\pm$0.1 &               &22.84  &  19.2&  0.09   &  DR2 \\ 
 J112256.47+340641.3   &0.043& 0.07 & 43.7  &      &  3.6$\pm$0.1 &               &23.29  &  16.6&  0.43   &  DR2 \\ 
 J112625.19+520503.5   &0.048& 0.12 & 37.1  &      &  5.0$\pm$0.1 &               &23.31  &  7.8 &  0.70   &  DR2 \\ 
 J112727.52+400409.4   &0.035& 0.09 & 45.2  &      &  5.9$\pm$0.1 &               &23.12  & 14.8 &  0.50   &  DR2 \\ 
 J113449.29+490439.4   &0.033& 0.14 &125.5  &140.0 &  6.8$\pm$0.1 &Two-sided      &23.51  & 25.1 &  0.72   &  DR2 \\ 
 J113637.14+510008.5   &0.050& 0.10 &  2.0  &      &  3.8$\pm$0.3 &               &22.08  &  7.8 & -0.61   &  DR2 \\ 
 J114232.84+262919.9   &0.030& 0.11 &  8.6  &      &  3.3$\pm$0.2 &               &22.27  & 44.7 & -0.74   &   P  \\ 
 J114804.60+372638.0   &0.042& 0.06 &  5.3  &      &  5.0$\pm$0.2 &               &22.35  & 29.2 & -0.76   &  DR2 \\ 
 J115531.39+545200.4   &0.050& 0.33 & 21.0  &      &  2.5$\pm$0.2 &               &23.10  & 31.8 & -0.19   &  DR2 \\ 
 J120551.46+203119.0   &0.024& 0.21 &104.5  &      &  2.9$\pm$0.2 &               &23.16  &102.7 &  0.01   &   P  \\ 
 J120607.81+400902.6   &0.037& 0.05 &  8.0  &      &  2.8$\pm$0.1 &               &22.42  &  8.9 & -0.05   &  DR2 \\ 
 J121329.27+504429.4   &0.031& 0.27 &216.3  &      &  6.4$\pm$0.1 &               &23.70  &102.7 &  0.33   &  DR2 \\ 
 J121951.65+282521.3   &0.027& 0.10 &  7.6  &      &  8.0$\pm$0.4 &               &22.12  &  8.0 & -0.02   &  DR2 \\ 
 J122421.31+600641.2   &0.044& 0.08 & 16.0  &      &  5.4$\pm$0.1 &               &22.87  &  5.7 &  0.46   &  DR2 \\ 
 J123011.85+470022.7   &0.039& 0.16 &118.6  &      &  3.3$\pm$0.1 &               &23.64  & 87.5 &  0.14   &  DR2 \\ 
 J130837.91+434415.1   &0.036& 0.14 &131.3  &      &  5.1$\pm$0.1 &               &23.61  & 52.3 &  0.41   &  DR2 \\ 
 J133042.51+323249.0   &0.034& 0.09 & 19.9  &      &  4.1$\pm$0.1 &               &22.74  & 16.9 &  0.07   &  DR2 \\ 
 J133737.49+155820.0   &0.026& 0.76 & 50.8  &      &  7.7$\pm$0.4 &               &22.92  & 30.1 &  0.23   &   P  \\ 
 J134159.72+294653.5   &0.045& 0.09 & 14.9  &      &  4.9$\pm$0.1 &               &22.86  &  9.4 &  0.21   &  DR2 \\ 
 J135036.01+334217.3   &0.014& 0.09 &352.0  &      &  3.9$\pm$0.1 &               &23.22  & 99.7 &  0.56   &  DR2 \\ 
 J140528.32+304602.0   &0.025& 0.08 &  2.4  &      &  5.0$\pm$0.5 &               &21.56  &  7.6 & -0.52   &  DR2 \\ 
 J142724.23+372817.0   &0.032& 0.08 & 52.7  &      &  7.2$\pm$0.1 &               &23.11  & 10.8 &  0.71   &  DR2 \\ 
 J143312.96+525747.3   &0.047& 0.09 &  8.6  &      &  5.1$\pm$0.1 &               &22.66  & 15.2 & -0.25   &  DR2 \\ 
 J143424.79+024756.2   &0.028& 0.36 &  6.3  &      &  5.5$\pm$1.2 &               &22.07  &  9.0 & -0.16   &   P  \\ 
 J143620.38+051951.5   &0.029& 0.48 & 15.6  &      &  6.7$\pm$0.8 &               &22.50  & 26.7 & -0.24   &   P  \\ 
 J152010.94+254319.3   &0.034& 0.16 & 10.7  &      &  5.6$\pm$0.5 &               &22.47  & 17.1 & -0.21   &   P  \\ 
 J152151.85+074231.7   &0.044& 0.64 & 31.4  &      &  6.0$\pm$0.2 &               &23.16  &  7.7 &  0.63   &   P  \\ 
 J153016.15+270551.0   &0.033& 0.15 & 18.9  &      &  3.5$\pm$0.3 &               &22.69  & 13.4 &  0.15   &   P  \\ 
 J154147.28+453321.7   &0.037& 0.10 & 14.8  & 30.9 &  9.0$\pm$0.1 &Head-tail      &22.69  &  6.2 &  0.39   &  DR2 \\ 
 J154426.93+470024.2   &0.038& 0.09 & 32.9  &      &  2.2$\pm$0.1 &               &23.06  & 17.4 &  0.29   &  DR2 \\ 
 J154451.23+433050.6   &0.037& 0.15 &  3.2  &      &  5.3$\pm$0.3 &               &22.02  & 12.5 & -0.61   &  DR2 \\ 
 J155951.61+255626.3   &0.045& 0.62 & 82.5  &      &  6.9$\pm$0.4 &               &23.60  & 29.2 &  0.46   &   P  \\ 
 J155953.99+444232.4   &0.042& 0.12 &221.7  &      &  5.1$\pm$0.1 &               &23.97  & 58.8 &  0.59   &  DR2 \\ 
 J160426.51+174431.1   &0.041& 0.38 & 49.4  &110.9 &  7.7$\pm$0.7 & Head-tail     &23.30  & 71.8 & -0.17   &   P  \\ 
 J160523.84+143851.6   &0.041& 0.21 & 24.4  & 26.4 & 10.4$\pm$0.5 & Core-jet      &22.99  &  7.5 &  0.53   &   P  \\ 
 J160641.83+084436.8   &0.047& 0.29 &  9.4  &      &  6.3$\pm$0.6 &               &22.69  &  8.8 &  0.03   &   P  \\ 
 J161238.84+293836.9   &0.032& 0.09 &  5.8  &      &  3.8$\pm$0.2 &               &22.15  & 22.8 & -0.61   &  DR2 \\ 
 J161256.85+095201.5   &0.017& 0.33 & 38.2  &      &  7.3$\pm$0.6 &               &22.42  & 21.3 &  0.26   &   P  \\ 
 J162944.98+404841.6   &0.029& 0.07 & 14.4  &      &  6.2$\pm$0.1 &               &22.46  &  6.9 &  0.33   &  DR2 \\ 
 J164925.86+360321.3   &0.032& 0.08 & 42.6  &      &  5.0$\pm$0.1 &               &23.02  & 10.5 &  0.63   &  DR2 \\ 
 J165830.05+252324.9   &0.033& 0.22 &  4.5  &      &  7.2$\pm$0.7 &               &22.07  & 14.8 & -0.53   &   P  \\ 
 J170358.49+241039.5   &0.031& 0.22 & 45.8  & 58.7 & 12.5$\pm$0.7 & Diffuse       &23.02  & 22.8 &  0.31   &   P  \\ 
 J172215.41+304239.8   &0.046& 0.10 & 43.6  & 63.2 & 24.0$\pm$0.2 & Two-sided     &23.34  &  5.7 &  0.91   &  DR2 \\ 

\hline
\end{longtable}

\smallskip
\small{Column description: (1) name; (2) redshift; (3) r.m.s. of the
  LOFAR images in mJy beam$^{-1}$; (4) flux densities (in mJy) at 150
  MHz of the central radio component. Errors are dominated by the
  $\sim$10\% flux scale uncertainty; (5) total 150 MHz flux density
  (in mJy) for the extended sources; (6) deconvolved major axis of the
  central source in arcseconds; (7) morphological description of the
  extended emission, when present; (8) logarithm of the luminosity (in
  $\WHz$) of the central component at 150 MHz; (9) flux density (in
  mJy) at 1.4 GHz from FIRST; (10) spectral index $\alpha$ between 150
  MHz and 1.4 GHz defined as $F_{\nu}\propto\,\nu^{-\alpha}$; errors
  for $\alpha$ are between 0.05 and 0.12; (11) image type: ``DR2'' =
  image in the DR2 area, ``P'' = individual LoTSS pointing with offset
  $<3^\circ$.}
\twocolumn

\end{appendix}
\end{document}